\documentclass[a4paper,12pt]{article}
\usepackage{graphicx}
\begin{document}
\title{Equation of state of strongly coupled\\ 
Hamiltonian lattice QCD at finite density}
\author{Yasuo Umino\\
\\
ECT$^*$\\
Strada delle Tabarelle 286\\
I--38050 Villazzano (Trento), Italy\\
and\\
Instituto de F\'{\i}sica Corpuscular -- C.S.I.C. \\
Departamento de F\'{\i}sica Te\`orica, Universitat de Val\`encia \\
E--46100 Burjassot, Val\`encia, Spain}
\date{\today}
\maketitle
\begin{abstract}
We calculate the equation of state of strongly coupled Hamiltonian lattice QCD at finite 
density by constructing a solution to the equation of motion corresponding to an effective 
Hamiltonian using Wilson fermions. We find that up to and beyond
the chiral symmetry restoration density the pressure of the quark Fermi sea can be negative 
indicating its mechanical instability. This result is in qualitative agreement 
with continuum models and should be verifiable by future numerical simulations.

\vskip 0.25cm
\noindent Submitted to Physics Letters B
\end{abstract}
\vfill\eject
Simulating Quantum Chromodynamics (QCD) at finite density is one of the outstanding 
problems in lattice gauge theory \cite{cre00}. In fact, because of the sign probelm 
there are currently no reliable numerical simulations of finite density QCD with three 
colors even in the strong coupling limit \cite{alo00}. This is a rather frustrating 
situation in view of the current intense interest in finite density QCD fueled by the 
phenomenology of heavy ion collisions, neutron stars, early universe and color 
superconductivity. Therefore even a qualitative description of finite density lattice 
QCD is welcome. 

One method of studying finite density lattice QCD is to invoke the strong coupling 
approximation where analytical methods are applicable. Strongly coupled lattice QCD at 
finite quark chemical potential $\mu$ and temperature $T$ has previously been studied 
analytically both in the Euclidean \cite{dam85,ilg85,bil92} and in the Hamiltonian 
\cite{pat84,van84,ley88,gre00} formulations. One of the main objectives of these studies  
was to investigate the nature of chiral symmetry restoration at finite $T$ and/or $\mu$. 
This has been accomplished by constructing some effective action or Hamiltonian for strongly 
coupled lattice QCD using Kogut--Susskind fermions. Except for \cite{ley88} these effective 
descriptions involve composite meson and baryon fields which are treated in the mean field 
approximation.\footnote{The work of \cite{ley88} does not involve composite fields but 
the approach is equivalent to the mean field approximation.} The consensus is that at zero 
or low $T$, strongly coupled lattice QCD at finite $\mu$ undergoes a first order chiral phase 
transition from the broken symmetry phase below a critical chemical potential $\mu_C$ to a 
chirally symmetric phase above $\mu_C$.

In this letter we present a calculation of the equation of state of strongly coupled lattice 
QCD at finite density in the Hamiltonian formulation using Wilson fermions. We find that up 
to and beyond the chiral symmetry restoration density the pressure of the many body system 
can be negative indicating its mechanical instability. This new result is in qualitative 
agreement with those obtained using continuum effective QCD models at finite density 
\cite{bub96,alf98} and should be verifiable by future numerical simulations. 

As in previous studies on this subject we begin with an effective description of strongly  
coupled lattice QCD. We shall use Smit's effective Hamiltonian \cite{smi80} which involves 
only the quark field $\Psi$ with a nearest neighbour interaction. This effective Hamiltonian 
has been studied in free space by Smit \cite{smi80} and by Le~Yaouanc et al. \cite{ley86} who 
subsequently extended their analysis to finite $T$ and $\mu$ using Kogut--Susskind fermions 
\cite{ley88}. A similar effective Hamiltonian has recently been derived by 
Gregory et al. \cite{gre00} to study strongly coupled lattice QCD at finite $\mu$, again using 
Kogut--Susskind fermions. 

Henthforth we shall adopt the notation of Smit \cite{smi80}, set the lattice spacing to 
unity and work in momentum space. Then the charge conjugation symmetric form of Smit's 
Hamiltonian using Wilson fermions may be written as
\begin{eqnarray}
H_{\rm eff}
& = &
\frac{1}{2} M_0 \left( \gamma_0\right)_{\rho\nu} \sum_{\vec{p}} 
\biggl[ \bigl(\Psi^\dagger_{a\alpha}\bigr)_\rho(\vec{p}\,),
\bigl(\Psi_{a\alpha}\bigr)_\nu(-\vec{p}\,) \biggr]_-
\nonumber \\
&   & 
\!\!\!\!\!
-\frac{K}{8N_{\rm c}} \sum_{\vec{p}_1,\ldots,\vec{p}_4 }\sum_{l}
\delta_{\vec{p}_1+\cdot\cdot\cdot+\vec{p}_4,\vec{0}}
\Biggl[ e^{i((\vec{p}_1+\vec{p}_2)\cdot \hat{n}_l)} 
+ e^{i((\vec{p}_3+\vec{p}_4)\cdot \hat{n}_l)} \Biggr]
\nonumber \\
&   &
\!\!\!\!\!
\otimes \Biggl[
\left( \Sigma_l \right)_{\rho\nu} 
\bigl(\Psi^\dagger_{a\alpha}\bigr)_\rho(\vec{p}_1\,)  
\bigl( \Psi_{b\alpha}\bigr)_\nu(\vec{p}_2\,)
-
\left( \Sigma_l \right)^\dagger_{\rho\nu}
\bigl(\Psi_{a\alpha}\bigr)_\nu(\vec{p}_1\,)
\bigl( \Psi^\dagger_{b\alpha}\bigr)_\rho(\vec{p}_2\,) \Biggr]
\nonumber \\
&   &
\!\!\!\!\!
\otimes \Biggl[
\left( \Sigma_l \right)^\dagger_{\gamma\delta} 
\bigl(\Psi^\dagger_{b\beta}\bigr)_\gamma(\vec{p}_3\,)  
\bigl(\Psi_{a\beta}\bigr)_\delta(\vec{p}_4\,)
-
\left( \Sigma_l \right)_{\gamma\delta}
\bigl(\Psi_{b\beta}\bigr)_\delta(\vec{p}_3\,)
\bigl( \Psi^\dagger_{a\beta}\bigr)_\gamma(\vec{p}_4\,) \Biggr]
\label{eq:heffp}
\end{eqnarray}
where $\left(\Sigma_l \right) = -i \left( \gamma_0 \gamma_l - ir\gamma_0 \right)$ with
the Wilson parameter $r$ taking on values between 0 and 1. 
In the above Hamiltonian color, flavor and Dirac indices are denoted by $(a, b)$, 
$(\alpha, \beta)$ and $(\rho, \nu, \gamma, \delta)$, respectively, and summation
convention is implied. $N_c$ is the number of colors.

The three parameters in $H_{\rm eff}$ are the Wilson parameter $r$, the current quark mass 
$M_0$ and the effective coupling constant $K = 2N_c/(N_c^2-1)\,1/g^2$ where $g$ 
is the QCD coupling constant. When $r = M_0 = 0$ the Hamiltonian possesses a $U(4N_f)$ 
symmetry with $N_f$ being the number of flavors. This symmetry is spontaneously broken down to 
$U(2N_f)\otimes U(2N_f)$ accompanied by the appearance of $8N_f^2$ Goldstone bosons 
\cite{smi80}. A finite current quark mass also breaks the original $U(4N_f)$ symmetry, albeit 
explicitly, down to $(2N_f)\otimes U(2N_f)$. Introduction of a finite Wilson parameter 
further breaks the latter symmetry explicitly down to $U(N_f)$ thereby solving the fermion 
doubling problem.

The above Hamiltonian has been derived in the temporal gauge using second order degenerate 
perturbation theory, and provides an effective description of only the ground state of strongly 
coupled lattice QCD \cite{smi80}. This ground state is the one in which no links are excited 
by the color electric flux. In the strong coupling limit the energy of one excited color 
electric flux link is 
\begin{equation}
E = \frac{1}{2N_c}(N_c^2 - 1)\,g^2 = \frac{1}{K}
\label{eq:ELINK}
\end{equation}
Therefore an extension of $H_{\rm eff}$ to finite $T$ and/or $\mu$ will be valid as long as 
$T,\,\mu < 1/K$ \cite{ley88}.\footnote{Note that in \cite{ley88} $E$ has been approximated 
by $E \approx N_c\,g^2$.} We shall see that this condition is satisfied in the present 
work.

Our method for obtaining the equation of state of strongly coupled lattice QCD at finite 
density using $H_{\rm eff}$ does not involve composite fields. Instead we explicitly 
construct a solution to the equation of motion corresponding to $H_{\rm eff}$ for all 
densities and use it to calculate the equation of state. For free space such a solution 
has been found in \cite{umi00}. This solution has the same structure as the free lattice 
Dirac field and exactly diagonalizes $H_{\rm eff}$ to second order in field operators. 
It obeys the free lattice Dirac equation with a dynamical quark mass which is determined 
by solving a gap equation. 

Temporarily dropping color and flavor indices this solution is given by
\begin{equation}
\Psi_\nu(t,\vec{p}\,) = 
b(\vec{p}\,) \xi_\nu(\vec{p}\,)e^{-i\omega(\vec{p}\,) t}
+ d^\dagger(-\vec{p}\,)\eta_\nu(-\vec{p}\,)e^{+i\omega(\vec{p}\,) t}
\label{eq:psip}
\end{equation}
with $\nu$ denoting the Dirac index. The annihilation
operators for particles $b$ and anti--particles $d$ annihilate an
interacting vacuum state and obey the free fermion anti--commutation
relations. The properties of the spinors $\xi$ and $\eta$ are given in
\cite{umi00}. The equation of motion for a free lattice Dirac field fixes
the excitation energy $\omega(\vec{p}\,)$ to be
\begin{equation}
\omega(\vec{p}\,) = \left( \sum_l {\rm sin}^2 (\vec{p}\cdot\hat{n}_l)
+ M^2(\vec{p}\,) \right)^{1/2}
\label{eq:EXEN}
\end{equation}
where $M(\vec{p}\,)$ is the dynamical quark mass.

The extension of the method developed in \cite{umi00} to finite $T$ and $\mu$ is 
accomplished in two steps. 
The first one is to make the following trivial replacement of the current quark mass term 
in $H_{\rm eff}$ Eq.~(\ref{eq:heffp})
\begin{equation}
M_0 \left( \gamma_0\right)_{\rho\nu} \rightarrow 
M_0 \left( \gamma_0\right)_{\rho\nu} - \mu_0 \delta_{\rho\nu} 
\end{equation}
where $\mu_0$ is the quark chemical potential. Note that 
$\mu_0$ should {\em not} be identified with the total chemical potential $\mu_{\rm tot}$ 
of the interacting many body system. As we shall see below the interaction will induce a 
correction to $\mu_0$ which in general is momentum dependent. We shall therefore 
refer to $\mu_0$ as the "bare" quark chemical potential and treat it as a parameter.

The second step is to observe that the annihilation operators $b$ and $d$ in 
Eq.~(\ref{eq:psip}) no longer annihilate the interacting vacuum state at finite $T$ and 
$\mu$ denoted as $|\,{\cal G}(T,\mu)\rangle$. In order to construct operators that annihilate
$|\,{\cal G}(T,\mu)\rangle$ we apply a generalized thermal Bogoliubov transformation to the 
$b$ and $d$ operators following the formalism of thermal field dynamics \cite{umezawa}
\begin{eqnarray}
b(\vec{p}\:) 
& = & 
\alpha_p B(\vec{p}\:) - \beta_p \tilde{B}^{\dagger}(-\vec{p}\:)
\label{eq:BTb} \\
d(\vec{p}\:) 
& = &
\gamma_p D(\vec{p}\:) - \delta_p \tilde{D}^{\dagger}(-\vec{p}\:)
\label{eq:TBT}
\end{eqnarray}
The thermal field operators $B$ and $\tilde{B}^\dagger$
annihilate a quasi--particle and  create a quasi--hole at finite $T$ and
$\mu$, respectively, while $D$ and  $\tilde{D}^\dagger$ are the
annihilation operator for a quasi--anti--particle  and creation operator for
a quasi--anti--hole, respectively. 

These thermal annihilation operators
annihilate the interacting thermal vacuum state {\em for each $T$ and
$\mu$}.  
\begin{equation}
B(\vec{p}\:) |\, {\cal G}(T,\mu)\rangle 
=\tilde{B}(\vec{p}\:) |\, {\cal G}(T,\mu)\rangle 
= D(\vec{p}\:) |\, {\cal G}(T,\mu)\rangle 
=\tilde{D}(\vec{p}\:) |\, {\cal G}(T,\mu)\rangle 
= 0
\label{eq:AVAC}
\end{equation}
The thermal doubling of the Hilbert space accompanying the thermal
Bogoliubov transformation is implicit in Eq.~(\ref{eq:AVAC})
where the vacuum state which is annihilated by thermal operators 
$B$, $\tilde{B}$, $D$ and $\tilde{D}$ is defined. Since we
shall be working only in the space of quantum field operators it 
is not necessary to specify the structure of $|\, {\cal G}(T,\mu)\rangle$. 

The thermal operators also satisfy the fermion anti--commutation relations
\begin{eqnarray}
\delta_{\vec{p},\vec{q}}
& = &
\biggl[B^{\dagger}(\vec{p}\:), B(\vec{q}\:) \biggr]_+ =
\biggl[\tilde{B}^{\dagger}(\vec{p}\:), \tilde{B}(\vec{q}\:) \biggr]_+ 
\nonumber\\
& = &
\biggl[D^{\dagger}(\vec{p}\:), D(\vec{q}\:) \biggr]_+ = 
\biggl[\tilde{D}^{\dagger}(\vec{p}\:), \tilde{D}(\vec{q}\:) \biggr]_+ 
\label{eq:THOP}
\end{eqnarray}
with vanishing anti--commutators for the remaining combinations.
The coefficients of the transformation are $\alpha_p = \sqrt{1-n_p^-}$,
$\beta_p = \sqrt{n_p^-}$, $\gamma_p = \sqrt{1-n_p^+}$ and
$\delta_p = \sqrt{n_p^+}$, where 
$n_p^{\pm}= [e^{(\omega_p \pm \mu)/(k_B T)}+1]^{-1}$ are the Fermi 
distribution functions for particles and anti--particles.
They are chosen so that the total particle number densities are given by
\begin{eqnarray}
n_p^- 
& = &
\langle\, {\cal G}(T,\mu)|\, b^{\dagger}(\vec{p}\:)b(\vec{p}\:)
|\, {\cal G}(T,\mu)\rangle\\
n_p^+
& = &
\langle\, {\cal G}(T,\mu)|\, d^{\dagger}(\vec{p}\:)d(\vec{p}\:)
|\, {\cal G}(T,\mu)\rangle %
\label{eq:DIST}
\end{eqnarray}
Hence in this approach temperature and chemical
potential are introduced simultaneously through the coefficients of the
thermal Bogoliubov transformation and are treated on an equal
footing. We stress that the chemical potential appearing in the
Fermi distribution functions is the {\em total} chemical potential of the
interacting many body system.

In addition to these changes, we demand that our ansatz satisfies the
equation of motion corresponding to the free lattice Dirac Hamiltonian with a
chemical potential term given by
\begin{eqnarray}
H^0 
& = &
\frac{1}{2} \sum_{\vec{p}}
\Biggl[ - \sum_l {\rm sin}(\vec{p}\cdot\hat{n}_l)(\gamma_0\gamma_l)_{\rho\nu}
+ M(\vec{p}\,)(\gamma_0)_{\rho\nu} - \mu_{\rm tot} \delta_{\rho\nu} \Biggr]
\nonumber\\
&   &
\;\;\;\;\;\;\;\;\;\;\;\;\;\;\;\;\;\;\;\;\;\;\;\;\;\;\;\;\;\;\;\;\;\;\;\;\;\;\;\;
\;\;\;\;\;\;\;\;\;\;
\otimes
\Bigg[ \Psi_\rho^\dagger(t, \vec{p}\,), \Psi_\nu(t, \vec{p}\,) \Biggr]_-
\label{eq:H0}
\end{eqnarray}
As in \cite{umi00} the mass $M(\vec{p}\,)$ is identified with the dynamical quark 
mass. Thus our ansatz at finite $T$ and $\mu$ is 
\begin{eqnarray}
\Psi_\nu(t, \vec{p}\:) 
& = &
\biggl[ \alpha_p B(\vec{p}\:) - \beta_p \tilde{B}^{\dagger}(-\vec{p}\:)\biggr]
\xi_\nu(\vec{p}\,)e^{-i[\omega(\vec{p}\,)-\mu_{\rm tot}] t}  \nonumber \\
&    &
\;\;\;\;\;\;\;\;\;\;\; +
\biggl[ \gamma_p D^{\dagger}(-\vec{p}\:) - \delta_p
\tilde{D}(\vec{p}\:) \biggr] \eta_\nu(-\vec{p}\,)e^{+i[\omega(\vec{p}\,) +
\mu_{\rm tot}] t}
\label{eq:HAAGTMU}
\end{eqnarray}
The spinors $\xi$ and $\eta$ obey the same properties as
in free space and the excitation energy $\omega(\vec{p}\,)$ has the same
form as in Eq.~(\ref{eq:EXEN}). The unknown quantities in Eq.~(\ref{eq:HAAGTMU}) are
the dynamical quark mass and the total chemical potential.

In this work we shall take the $T \rightarrow 0$ limit which amounts to setting 
$\gamma_p = 1$ and $\delta_p = 0$ in Eq.~(\ref{eq:TBT}) thereby suppressing the excitation of
anti--holes. In this limit $\beta_p^2$ becomes the Heaviside function 
$\beta_p^2 = \theta( \mu_{\rm tot} - \omega(\vec{p}\,))$ defining the Fermi momentum 
$\vec{p}_F$ where 
\begin{equation}
\mu_{\rm tot} = \left( \sum_l {\rm sin}^2 (\vec{p}_F\cdot\hat{n}_l) 
+ M^2(\vec{p}_F)\right)^{1/2}
\end{equation}
One of the simplest quantities to calculate using the ansatz of Eq.~(\ref{eq:HAAGTMU}) in 
the $T \rightarrow 0$ limit is the quark number density $n$ given by
\begin{equation}
n = \frac{1}{2 V N_f N_c}\langle\bar{\Psi}\gamma_0\Psi\rangle = 
\sum_{\vec{p}} \theta( \mu_{\rm tot} - \omega(\vec{p}\,)) 
\end{equation}
Therefore, above a sufficiently large value of $\mu_{\rm tot}$ the quark number density 
becomes a constant which with the present normalization will equal unity. This saturation 
effect is purely a lattice artifact originating from the ${\rm sin}^2 (\vec{p}\cdot\hat{n}_l)$ 
term in $\omega(\vec{p}\,)$.  

Another quantity that may be readily calculated using the $T \rightarrow 0$ ansatz is the
chiral condensate. It is found to be proprotional to the dynamical quark mass
\begin{equation}
\frac{1}{2V N_f N_c }\langle \bar{\Psi}\Psi \rangle = 
-\sum_{\vec{p}} \alpha_p^2 \frac{M(\vec{p}\,)}{\omega(\vec{p}\,)}
\label{eq:cond}
\end{equation}
Below we shall derive a gap equation for $M(\vec{p}\,)$ and show that for a given 
physically reasonable set of 
parameters there exists a critical chemical potential above which $M(\vec{p}\,) = 0$.
Thus the chiral condensate may be identified as being the order parameter for the chiral 
phase transition at finite density.

However before deriving the gap equation we shall demonstrate that in the $T \rightarrow 0$ 
limit the ansatz shown in Eq.~(\ref{eq:HAAGTMU}) exactly diagonalizes the effective Hamiltonian 
to second order in field operators for all densities. We make use of the 
fact that our ansatz satisfies the equation of motion corresponding to the free lattice Dirac 
Hamiltonian $H^0$ given in Eq.~(\ref{eq:H0}). Therefore we have the relation
\begin{equation}
:\Bigl[ \bigl(\Psi_{a\alpha}\bigr)_\mu(t, \vec{q}\,), H^0 \Bigr]_- :
\;\;\;\;
=
\;\;\;\;
:\Bigl[ \bigl(\Psi_{a\alpha}\bigr)_\mu(t, \vec{q}\,), H_{\rm eff} \Bigr]_- :
\label{eq:CRUX}
\end{equation}
where the symbol :  : denotes normal ordering with respect
to the vacuum at zero temperature $|\, {\cal G}(T=0,\mu)\rangle$. Evaluating both sides of
Eq.~(\ref{eq:CRUX}) we obtain
\begin{eqnarray}
\lefteqn{\!\!\!\!\!\!\!
\Bigl[ \sum_l {\rm sin}(\vec{q}\cdot\hat{n}_l)(\gamma_0\gamma_l)_{\rho\delta}
+ M(\vec{q}\,)(\gamma_0)_{\rho\delta} - \mu_{\rm tot}\delta_{\rho\delta} \Bigr]
\bigl(\Psi_{a\alpha}\bigr)_\delta(t, \vec{q}\,) =
} \nonumber \\
&   &
\Biggl\{
M_0 \left(\gamma_0\right)_{\rho\delta} - \mu_0 \delta_{\rho\delta}
\nonumber \\
&   & 
+ \frac{1}{N_{\rm c}}K\sum_{\vec{p}}\sum_l 
\alpha_p^2 \Lambda^+_{\nu\gamma}(\vec{p}\,)
\nonumber \\
&   &
\otimes\left[ \cos\left( \vec{p}-\vec{q}\;\right)\cdot\hat{n}_l
\Biggl(
\bigl(\Sigma_l\bigr)_{\gamma\nu}\bigl(\Sigma_l\bigr)^\dagger_{\rho\delta}
+
\bigl(\Sigma_l\bigr)^\dagger_{\rho\nu}\bigl(\Sigma_l\bigr)_{\gamma\delta}
\Biggr)\right.
\nonumber \\
&   &
\:\:\:
\left.
+
\cos\left( \vec{p}+\vec{q}\;\right)\cdot\hat{n}_l
\Biggl(
\bigl(\Sigma_l\bigr)^\dagger_{\gamma\nu}\bigl(\Sigma_l\bigr)^\dagger_{\rho\delta}
+
\bigl(\Sigma_l\bigr)_{\rho\nu}\bigl(\Sigma_l\bigr)_{\gamma\delta}
\Biggr)\right]
\nonumber \\
&   &
-\frac{1}{N_{\rm c}}\frac{K}{4} \sum_{\vec{p},\vec{q}} \sum_l 
\left[ 2\alpha_p^2 \Lambda^+_{\nu\gamma}(\vec{p}\,) - \delta_{\nu\gamma} \right]
\nonumber \\
&   &
\otimes\left[ N_{\rm c}
\Biggl(
\bigl(\Sigma_l\bigr)_{\rho\nu}\bigl(\Sigma_l\bigr)^\dagger_{\gamma\delta}
+
\bigl(\Sigma_l\bigr)^\dagger_{\rho\nu}\bigl(\Sigma_l\bigr)_{\gamma\delta}
\Biggr)\right.
\nonumber \\
&   &
\:\:\:
\left.
+
\cos\left( \vec{p}+\vec{q}\;\right)\cdot\hat{n}_l
\Biggl(
\bigl(\Sigma_l\bigr)^\dagger_{\rho\nu}\bigl(\Sigma_l\bigr)^\dagger_{\gamma\delta}
+
\bigl(\Sigma_l\bigr)_{\rho\nu}\bigl(\Sigma_l\bigr)_{\gamma\delta}
\Biggr)\right] \Biggr\} \bigl(\Psi_{a\alpha}\bigr)_\delta(t, \vec{q}\,)
\label{eq:EOM1}
\end{eqnarray}
with $\Lambda^+(\vec{p}\,) \equiv  \xi(\vec{p}\,)\otimes\xi^\dagger(\vec{p}\,)$
being the positive energy projection operator defined in \cite{umi00}.

To second order in field operators the off--diagonal Hamiltonian is given by
\begin{eqnarray}
H_{\rm off}|\, {\cal G}(0, \mu)\rangle
& = &
\sum_{\vec{q}} \Biggl\{ \alpha_q \xi^\dagger_\rho(\vec{q}\,)
\left[ M_0\bigl( \gamma_0 \bigr)_{\rho\delta} - \mu_0 \delta_{\rho\delta} \right]
\nonumber \\
&   &
+ \frac{1}{N_{\rm c}}K\sum_{\vec{p},\vec{q}}\sum_l \alpha_p^2 \alpha_q
\Lambda^+_{\nu\rho}(\vec{p}\,)
\nonumber \\
&   &
\otimes\, \xi^\dagger_\gamma(\vec{q}\,)
\left[ \cos\left( \vec{p}-\vec{q}\;\right)\cdot\hat{n}_l
\Biggl(
\bigl(\Sigma_l\bigr)_{\rho\nu}\bigl(\Sigma_l\bigr)^\dagger_{\gamma\delta}
+
\bigl(\Sigma_l\bigr)^\dagger_{\rho\nu}\bigl(\Sigma_l\bigr)_{\gamma\delta}
\Biggr)\right.
\nonumber \\
&   &
\left.
+
\cos\left( \vec{p}+\vec{q}\;\right)\cdot\hat{n}_l
\Biggl(
\bigl(\Sigma_l\bigr)^\dagger_{\rho\nu}\bigl(\Sigma_l\bigr)^\dagger_{\gamma\delta}
+
\bigl(\Sigma_l\bigr)_{\rho\nu}\bigl(\Sigma_l\bigr)_{\gamma\delta}
\Biggr)\right]  
\nonumber \\
&   &
-\frac{1}{N_{\rm c}}\frac{K}{4} \sum_{\vec{p},\vec{q}} \sum_l \alpha_q
\left[ 2\alpha_p^2 \Lambda^+_{\nu\gamma}(\vec{p}\,) - \delta_{\nu\gamma} \right]
\nonumber \\
&   &
\otimes\,\xi^\dagger_\rho(\vec{q}\,)\left[ N_{\rm c}
\Biggl(
\bigl(\Sigma_l\bigr)_{\rho\nu}\bigl(\Sigma_l\bigr)^\dagger_{\gamma\delta}
+
\bigl(\Sigma_l\bigr)^\dagger_{\rho\nu}\bigl(\Sigma_l\bigr)_{\gamma\delta}
\Biggr)\right.
\nonumber \\
&   &
\left.
+
\cos\left( \vec{p}+\vec{q}\;\right)\cdot\hat{n}_l
\Biggl(
\bigl(\Sigma_l\bigr)^\dagger_{\rho\nu}\bigl(\Sigma_l\bigr)^\dagger_{\gamma\delta}
+
\bigl(\Sigma_l\bigr)_{\rho\nu}\bigl(\Sigma_l\bigr)_{\gamma\delta}
\Biggr)\right] \Biggr\} \eta_\delta(-\vec{q}\,)
\nonumber \\
&   &
\otimes B^\dagger_{\alpha,a}(\vec{q}\,)D^\dagger_{\alpha,a}(-\vec{q}\,)
|\, {\cal G}(0, \mu)\rangle
\label{eq:hoff2}
\end{eqnarray}
From Eq.~(\ref{eq:hoff2}) we see that the elementary excitations of the effective
Hamiltonian are color singlet (quasi) quark--anti--quark pairs coupled
to zero total three momentum. With the use of Eq.~(\ref{eq:EOM1}), the equation
of motion for the $\eta$ spinor and the orthonormality condition
$\xi_\nu^\dagger(\vec{p}\,)\eta_\nu(-\vec{p}\,) = 0$ \cite{umi00} we can show
that
\begin{eqnarray}
H_{\rm off}|\, {\cal G}(0, \mu)\rangle
& = &
\sum_{\vec{q}}
\Biggl\{ \alpha_q \xi^\dagger_\nu(\vec{q}\,)
\Bigl[ -\sum_l {\rm sin}(\vec{q}\cdot\hat{n}_l)(\gamma_0\gamma_l)_{\nu\delta}
\nonumber \\
&   &
\:\:\:\:\:\:\:\:\:\:\:\:\:\:\:\:\:
- M(\vec{q}\,)(\gamma_0)_{\nu\delta} + \mu_{\rm tot}\delta_{\nu\delta} \Bigr]
\eta_\delta(-\vec{q}\,) \Biggr\} 
\nonumber \\
&   &
\:\:\:\:\:\:\:\:\:\:\:\:\:\:\:\:\:\:\:\:\:\:\:\:\:\:\:\:\:\:\:\:\:\:\:\:\:\:\:\:
\otimes B^\dagger_{\alpha,a}(\vec{q}\,)D^\dagger_{\alpha,a}(-\vec{q}\,)
|\, {\cal G}(0, \mu)\rangle
\nonumber \\
& = &
\sum_{\vec{q}} \Biggl\{
\alpha_q \xi^\dagger_\nu(\vec{q}\,)
\Bigl[\omega(\vec{q}\,) + \mu_{\rm tot}\Bigr]\eta_\nu(-\vec{q}\,)
\Biggr\}
\nonumber \\
&   &
\:\:\:\:\:\:\:\:\:\:\:\:\:\:\:\:\:\:\:\:\:\:\:\:\:\:\:\:\:\:\:\:\:\:\:\:\:\:\:\:
\otimes B^\dagger_{\alpha,a}(\vec{q}\,)D^\dagger_{\alpha,a}(-\vec{q}\,)
|\, {\cal G}(0, \mu)\rangle
\nonumber \\
& = &
0
\end{eqnarray}
Therefore our ansatz exactly diagonalizes the effective
Hamiltonian to second order in field operators for all densities.

We now derive the equations for the dynamical quark mass and the total chemical 
potential and solve them to determine our solution Eq.~(\ref{eq:HAAGTMU}) for each
density. To accomplish this we explicitly evaluate the right hand side of 
Eq.~(\ref{eq:EOM1}) to reveal its Dirac structure. The result may be cast in the 
following compact form
\begin{eqnarray}
\lefteqn{ \!\!\!\!\!\!\!\!\!\!\!\!\!\!\!
\Bigl[ \sum_l \sin(\vec{q}\cdot\hat{n}_l)(\gamma_0\gamma_l)_{\nu\delta}
+ M(\vec{q}\,)(\gamma_0)_{\nu\delta} - \mu_{\rm tot}\delta_{\nu\delta} \Bigr]
\bigl(\Psi_{a\alpha}\bigr)_\delta(t, \vec{q}\,) =
} \nonumber \\
& &
\;\;\;\;\;\;\;\;\;\;\;
\Bigl[ A(\vec{q}\,)(\gamma_0\gamma_l)_{\nu\delta} +
B(\vec{q}\,)(\gamma_0)_{\nu\delta} + C(\vec{q}\,)\delta_{\nu\delta} \Bigr]
\bigl(\Psi_{a\alpha}\bigr)_\delta(t, \vec{q}\,)
\label{eq:EOM}
\end{eqnarray}
The equations for $M(\vec{p}\,)$ and $\mu_{\rm tot}$ are obtained by equating the 
coefficents of the $\gamma_0$ operator and the Kronecker delta function, respectively. 

The gap equation determining $M(\vec{p}\,)$ is given by the coefficient $B(\vec{q}\,)$
\begin{eqnarray}
M(\vec{q}\,)
& = &
B(\vec{q}\,)
\nonumber \\
& = &
M_0 + \frac{3}{2}K(1 - r^2) \sum_{\vec{p}} \left( 1 - \beta_p^2 \right)
\frac{M(\vec{p}\,)}{\omega(\vec{p}\,)}
\nonumber \\
&   &
+ \frac{K}{N_{\rm c}} \sum_{\vec{p},l} \left( 1 - \beta_p^2 \right)
\frac{M(\vec{p}\,)}{\omega(\vec{p}\,)}
\Biggl\{
8r^2  \cos(\vec{p}\cdot\hat{n}_l) \cos(\vec{q}\cdot\hat{n}_l)
\nonumber \\
&   &
\;\;\;\;\;\;\;\;\;\;\;\;\;\;
-\frac{1}{2}(1+r^2) \cos(\vec{p}+\vec{q}\, )\cdot\hat{n}_l
\Biggr\}
\label{eq:GAPEQ}
\end{eqnarray}
The structure of this gap equation is very similar to the one 
in free space ($\beta^2_p = 0$) found in \cite{umi00}. The dynamical quark
mass is a constant to lowest order in $N_c$ but becomes momentum dependent
once $1/N_c$ correction is taken into account. 

Similarly, the total chemical potential is 
given by the coefficient $C(\vec{q}\,)$ 
\begin{eqnarray}
\mu_{\rm tot}
& = &
- C(\vec{q}\,) \nonumber \\
& = &
\mu_0 + \frac{1}{4} \frac{K}{N_c} \sum_l\sum_{\vec{p}} \beta_p^2
\Bigl[ 2N_c \left( 1+r^2 \right) - 2 \left( 1-r^2 \right)
{\rm cos}\left( \vec{p} + \vec{q}\, \right) \Bigr]
\label{eq:MUTOT}
\end{eqnarray} 
Thus $\mu_{\rm tot}$ is a sum of the bare chemical potential $\mu_0$ and an interaction 
induced chemical potential which is proportional to the effective coupling constant $K$. 
Furthermore, the latter contribution 
to $\mu_{\rm tot}$ is momentum dependent and this dependence is a $1/N_c$ correction just as 
in the case of the gap equation. It should be 
noted that the above shifting of the bare chemical 
potential by the interaction is not a new effect. For example, in the well--known and 
well--studied Nambu--Jona--Lasinio model \cite{nam61} at finite $T$ and $\mu$ the interaction 
induces a contribution to the total chemical potential which is proportional to the number 
density \cite{asa89,kle92}. 

The two equations (\ref{eq:GAPEQ}) and (\ref{eq:MUTOT}) are coupled and therefore solutions 
for $M$ and $\mu_{\rm tot}$ must be found self--consistently for each value of the 
input parameter $\mu_0$. In Figure~1 we present $M$ as a function of $\mu_{\rm tot}$ 
for two values of $K$ determined by solving Eqs.~(\ref{eq:GAPEQ}) and (\ref{eq:MUTOT}) to 
${\cal O}(N_c^0)$, which is the same order in the $1/N_c$ expansion used to obtain results in 
all previous studies of strongly coupled lattice QCD. At this order in $N_c$ both the 
dynamical mass and the total chemical potential are momentum independent. The values of input 
parameters are $M_0 = 0$, $r = 0.25$ and $N_c$ =3. 

%
%
\begin{figure}[tbp]
\begin{center}
\includegraphics[height=10.5cm,width=6.5cm,angle=-90]{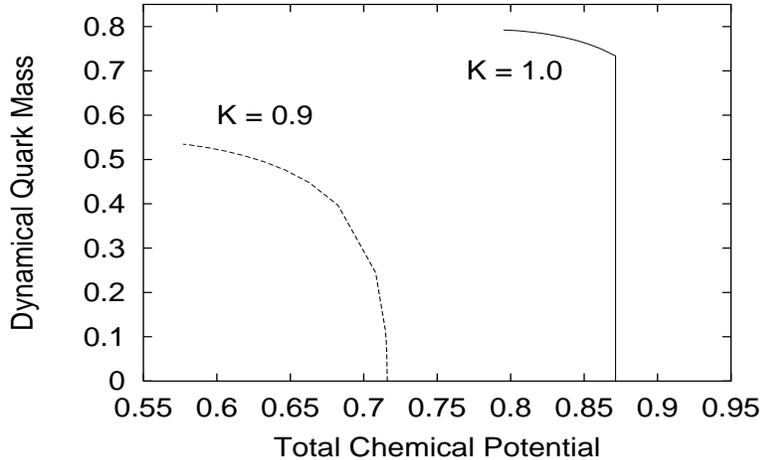} 
\end{center}
\caption{Dynamical quark mass $M$ as a function of total chemical potential $\mu_{\rm tot}$ 
for two values of effective coupling constant $K$. These results were obtained by solving 
Eqs.~(\ref{eq:GAPEQ}) and (\ref{eq:MUTOT}) self--consistently to lowest order in $N_c$ using 
$M_0 = 0$, $r = 0.25$ and $N_c$ = 3. There is a second order chiral phase transition when 
the effective coupling constant $K$ is 0.9. The order of the phase transition 
changes to first order when $K$ is increased to 1.0.}
\label{fig1}
\end{figure}

From the figure we see that the chiral phase transition can be either first or second order 
depending on the value of the effective coupling constant. When $K$ = 0.9 we find a second 
order phase transition with a critical chemical potential of 
$(\mu_{\rm tot})_C \approx 0.716$, while if the coupling constant is 
increased to $K$ = 1.0 the phase transition becomes first order with a larger critical chemical 
potential of $(\mu_{\rm tot})_C \approx 0.871$. Furthermore, we find that when $K = 0.9$ 
lattice saturation sets in around $\mu_{\rm tot} \approx 0.898$ while 
this effect takes place immediately above $(\mu_{\rm tot})_C$ for $K = 1.0$. These values of 
chemical potentials are smaller than the energy $E=1/K$ required to 
excite one color electric flux link as given in Eq.~(\ref{eq:ELINK}). Therefore with a 
reasonable set of parameters it is possible to extend Smit's effective Hamiltonian to finite 
density as was first pointed out in \cite{ley88}. 

Having solved for the dynamical quark mass and the total chemical potential 
we have constructed a solution 
to the equation of motion for $H_{\rm eff}$ in the $T \rightarrow 0$ limit to lowest order 
in $N_c$. In addition we have shown that this solution exactly diagonalizes the effective 
Hamiltonian to second order in field operators for all densities. Therefore we may use it to 
evaluate the vacuum energy density which to lowest order in $N_c$ is given by
\begin{eqnarray}
\lefteqn{
\frac{1}{V} \langle\, {\cal G}(0, \mu)\,| H_{\rm eff} |\, {\cal G}(0, \mu) \rangle =}
\nonumber \\
& &
\!\!\!\!\!\!\!\!\!\!
-2 N_c \sum_{\vec{p}} \Biggl\{
\alpha_p^2 \Biggl[ \frac{3}{2} K (1 + r^2) + \omega(\vec{p}\,)
+ \frac{M}{\omega(\vec{p}\,)} M_0 
- \mu_{\rm tot} \Biggr]
+ (1 + \beta_p^2) \mu_0 \Biggr\}
\label{eq:VED}
\end{eqnarray}
Numerically we find that the difference of the vacuum energy densities in the symmetry 
restored ($M = 0$) and broken ($M \neq 0$) phases of the theory is positive
\begin{equation}
\frac{1}{V}\langle\, {\cal G}(0, \mu)\,| H_{\rm eff} |\, {\cal G}(0, \mu) \rangle
|_{M = 0}
- \frac{1}{V}\langle\, {\cal G}(0, \mu)\,| H_{\rm eff} |\, {\cal G}(0, \mu) \rangle
|_{M \neq 0}
> 0
\end{equation}
Therefore the phase with broken chiral symmetry is the energetically favored phase. 

The equation of state is obtained by numerically evaluating the thermodynamic potential 
density using Eq.~(\ref{eq:VED}). In Figure~2 we plot the pressure as a function of 
$\mu_{\rm tot}$ for $K$ = 0.9 and 1.0 with $M_0 = 0$, $r = 0.25$ and $N_c = 3$. In both 
cases we 
find that the pressure of the quark Fermi sea is negative and monotonically decreasing in 
the broken symmetry phase. For $K$ = 0.9 the pressure remains negative but increasing in 
the symmetry restored phase, at least until the lattice saturation point, and has a cusp 
where the two phases meet. Unfortunately, we can not make a definite quantitative statement 
on the behaviour of the pressure in the symmetry restored phase for $K = 1.0$ due to lattice 
saturation, except to mention that there is a discontinuity when going from one phase to 
another. However, we may conclude that up to and beyond the chiral symmetry restoration 
point the quark Fermi sea can have negative pressure and therefore can be mechanically 
unstable with an imaginary speed of sound.

%
%
\begin{figure}[tbp]
\begin{center}
\includegraphics[height=10.5cm,width=6.5cm,angle=-90]{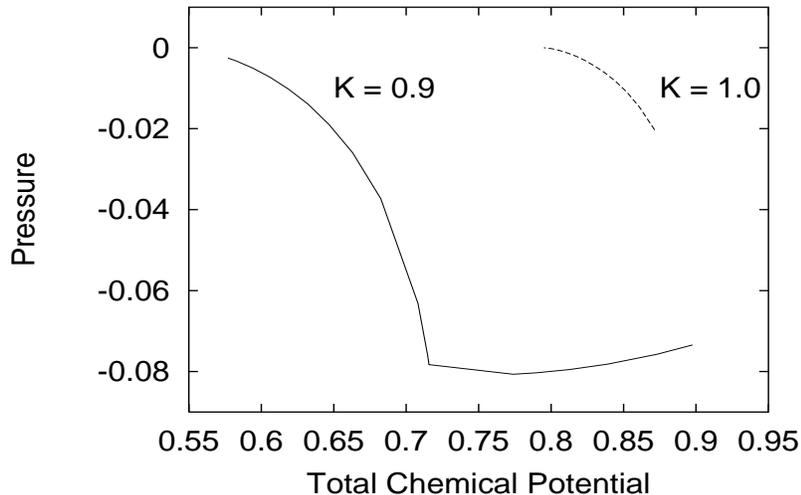} 
\end{center}
\caption{Pressure $P$ as a function total chemical potential $\mu_{\rm tot}$ for two 
values of effective coupling constant $K$. The results were obtained using the same 
parameters as in Figure~\ref{fig1}. For $K$ = 0.9, the critical chemical potential is 
$(\mu_{\rm tot})_C \approx 0.716$ and lattice saturation sets in around 
$\mu_{\rm tot} \approx 0.898$, while this effect takes
place right above $(\mu_{\rm tot})_C \approx 0.871$ for $K$ = 1.0.}
\label{fig2}
\end{figure}

Our conclusion regarding the (strongly coupled) quark matter stability at finite density 
is consistent with similar studies using the Nambu--Jona--Lasinio model \cite{bub96} and 
the effective instanton induced 't~Hooft interaction model \cite{alf98}. These mean
field calculations show that cold and dense quark matter may be unstable in the phase with 
spontaneously broken chiral symmetry, but can become stable in the symmetry restored phase 
at high enough density. In particular, the result for the pressure of cold and dense 
quark matter obtained in \cite{alf98} is qualitatively the same as the one shown in 
Figure~\ref{fig2}.\footnote{Compare Figure~1 of \cite{alf98} with Figure~\ref{fig2} of this
letter.} The possibility of unstable quark mattter lead the authors of 
\cite{bub96} and \cite{alf98} to speculate the 
formation of nucleon droplets, reminiscent of the MIT bag model, in the broken symmetry phase. 
We shall not dwell on such a speculation here since we are working in an artificial strong 
coupling regime. Nevertheless, it would be interesting to compare our result concerning the
negative pressure with future lattice simulations of finite density QCD at strong coupling.

In this work we studied the equation of state of strongly coupled lattice QCD in the 
Hamiltonian formulation using Wilson fermions. This was accomplished by constructing a
solution of the equation of motion correponding to an effective Hamiltonian which exactly 
diagonalizes the Hamiltonian to second order in field operators for all densities. We found 
that: the dynamical quark mass is in general momentum dependent; the interaction induces a 
momentum dependent contribution to the total chemical potential making it necessary 
to solve for the dynamical quark mass self--consistently with $\mu_{\rm tot}$; the elementary 
excitations of the theory consist of color singlet quark--anti--quark pairs coupled to zero 
total three momentum; and the broken symmetry phase is the energetically favored phase. 

To leading order in $N_c$ we find that the chiral phase transition can be either first or 
second order depending on the value of the effective coupling constant. In addition, the 
pressure of the 
strongly interacting many body system is found to be negative up to and beyond the chiral 
phase transition density. A similar behaviour for the pressure has been obtained with $r = 0$ 
which corresponds to using Kogut--Susskind fermions. Therefore our result concerning the 
negative pressure seems to be robust, at least to leading order in $N_c$, and should be 
verifiable by future numerical simulations of strongly coupled lattice QCD at finite density.

\vskip 2cm
\noindent{\bf Acknowledgements}
\vskip 1cm

I am indebted to M.--P.~Lombardo for comments and suggestions which lead to an
improvement of this manuscript, as well as to O.W.~Greenberg for reading 
the first draft. Part of this work was completed while I was at ECT$^*$ as a
Junior Visiting Scientist. I thank the Center for its hospitality and
generous support.

%

%
\end{document}